# PULSED SC PROTON LINAC

N. Ouchi, E. Chishiro, JAERI, Tokai, Japan
C. Tsukishima, K. Mukugi, MELCO, Kobe, Japan

*Abstract*

The superconducting (SC) proton linac is proposed in the JAERI/KEK Joint Project for a high-intensity proton accelerator in the energy region from 400 to 600 MeV. Highly stable fields in the SC cavities are required under the dynamic Lorentz force detuning. A new model describing the dynamic Lorentz detuning has been developed and the validity has been confirmed experimentally. The model has been applied successfully to the rf control simulation of the SC proton linac.

## 1 INTRODUCTION

The Japan Atomic Energy Research Institute (JAERI) and the High Energy Accelerator Research Organization (KEK) are proposing the Joint Project for High Intensity Proton Accelerator[1,2]. The accelerator consists of 600 MeV linac, 3 GeV RCS (Rapid Cycling Synchrotron) and 50 GeV synchrotron. SC structures are applied in the high energy part of the linac from 400 to 600 MeV. Momentum spread of the linac beams less than ±0.1% is required for the injection to the RCS. At the commissioning of the accelerator, 400 MeV beams will be injected into the RCS. In this period, the SC linac will provide the beams to the R&D for the ADS (Accelerator Driven System) and the machine study will be carried out to obtain acceptable beam quality for the RCS. In order to increase the beam intensity, the 600 MeV beams will be injected into the RCS after the machine study.

The linac accelerates $H^-$ beams in a pulsed operation; repetition rate of 50 Hz, beam duration of 0.5 ms, peak current of 50 mA and intermediate duty factor of 54 % by chopping. To meet the requirement of the RCS, rf amplitude and phase errors of the accelerating cavities should be less than ±1% and ±1deg, respectively. In the case of the SC cavities, the Lorentz force of the pulsed rf field induces dynamic deformation and detuning of the cavity, which disturb the accelerating field stability.

A new model which describes the dynamic Lorentz force detuning has been established for the rf control simulation in the pulsed SC linac. The validity of the model has been confirmed experimentally. The model has been applied to the rf control simulation.

A new model, comparison between calculated and experimental results, dynamic Lorentz detuning for the multi-cell cavity and the rf control simulation are presented in this paper.

## 2 MODEL FOR DYNAMIC LORENTZ DETUNING

### 2.1 Stationary Lorentz Detuning

SC cavities are deformed by the Lorentz force of their own electromagnetic field. The Lorentz pressure (P) on the cavity wall is presented by the equation[3],

$$P = \frac{1}{4}\left(\mu_0 H^2 - \varepsilon_0 E^2\right) ,$$

where H and E are magnetic and electric field strength on the cavity surface. Since the cavity deformation is proportional to the Lorentz pressure, the detuning ($\Delta f$) is proportional to the square of the accelerating field ($E_{acc}$) by assuming linearity between the deformation and the detuning. In our Joint Project, two kind of 972 MHz cavities, $\beta_g$ (geometrical $\beta$ of the cavity) = 0.729 and 0.771, are designed between 400 and 600 MeV region[4]. The detuning constants k (= $-\Delta f/E_{acc}^2$) of these 7-cell cavities are 1.61 and 1.42 Hz/(MV/m)$^2$, respectively.

### 2.2 Lorentz Vibration Model

To simulate the rf control, the time-dependent cavity field and detuning have to be solved simultaneously, because these affect each other. For this purpose, a new model which describes dynamic Lorentz detuning, named Lorentz Vibration Model, has been developed.

The basic idea of the model is listed below.

- The dynamic motion of the cavity is expanded in terms of the mechanical modes. This method is known as “Modal Analysis”.
- Cavity deformation for each mechanical mode is converted to the partial detuning for each mode using frequency sensitivity data.
- Total detuning is obtained by summing up the partial detuning for each mode.

According to this basic idea, we have obtained the Lorentz Vibration Model as the following equations.

$$\frac{d^2 \Delta f_k}{dt^2} + \frac{\omega_{mk}}{Q_{mk}} \frac{d\Delta f_k}{dt} + \omega_{mk}^2 \Delta f_k = K_k \left(\frac{V_C}{V_0}\right)^2 ,$$

$$K_k = \frac{1}{m_k}\left\{\left(\frac{\vec{df}}{du}\right)\bullet\left(\vec{a}_k\right)\right\}\left\{\left(\vec{F}_0\right)\bullet\left(\vec{a}_k\right)\right\} ,$$

$$\Delta f = \sum_k \Delta f_k ,$$

*where*

$V_c$ : cavity voltage (V)

$k$ : mechanical vibration mode number

$\Delta f_k$ : partial detuning for k - th mode (Hz)

$\Delta f$ : total detuning (Hz)

$\omega_{\mathrm{m}k}$ : angular frequency for $k$ - th mechanical mode (rad / s)

$Q_{mk}$ : quality factor for $k$ - th mechanical mode

$m_k$ : generalized mass for $k$ – $th$ mechanical mode (kg)

$\vec{F}_0$ : Lorentz force vector at cavity voltage of $V_0$ (N)

$\vec{a}_k$ : eigenvector for $k$ – $th$ mechanical mode

$\vec{df/du}$ : frequency sensitivity vector for displacement (Hz / m)

$u$ : cavity wall displacement (m)

The inner products of (df/du)•($a_k$) and ($F_0$)•($a_k$) mean the detuning sensitivity of *k-th* mechanical mode and the Lorentz force contribution to the *k-th* mechanical mode, respectively. Parameters of $\omega_{mk}$, $m_k$ and ($a_k$) are obtained from the structural analysis code, ABAQUS. ($F_0$) and (df/du) are obtained from the SUPERFISH results.

## *2.3 Pulsed Operation in the Vertical Test*

In order to observe the dynamic Lorentz detuning experimentally, a pulsed operation was carried out in the vertical test of a single-cell 600MHz cavity of $\beta_g$=0.886. In the test, one side of the cavity flange was fixed to the cryostat and the other side was free. The measurement was made at 4.2 K. Unloaded and loaded quality factors of the cavity were ~$9\times10^8$ and ~$9\times10^7$, respectively.

Figure 1 shows the rf power control signal, which is proportional to the amplifier output power (max. 300W), and the surface peak field of the cavity ($E_{peak}$) in a pulse. Rise time, flat top and repetition rate were 60 ms, 100 ms and 0.76 Hz, respectively. The cavity was excited even between the pulses with very low field ($E_{peak}$~0.7 MV/m) in order to keep lock of a PLL (Phase Locked Loop) circuit. Dynamic Lorentz detuning was measured by taking an FM control signal of the PLL circuit through a low path filter of 1 kHz. The signal was accumulated for about 40 pulses and averaged to eliminate random noises.

Figure 2 shows the dynamic Lorentz detuning obtained in the test. Vibration of the detuning was observed at the flat top and decay of the pulse. Impulses at the beginning of the rise and both ends of the flat top were due to the responses of the PLL circuit. The frequency and the quality factor of the vibration were estimated to be 122 Hz and ~60, respectively, by analysing the waveform at the decay.

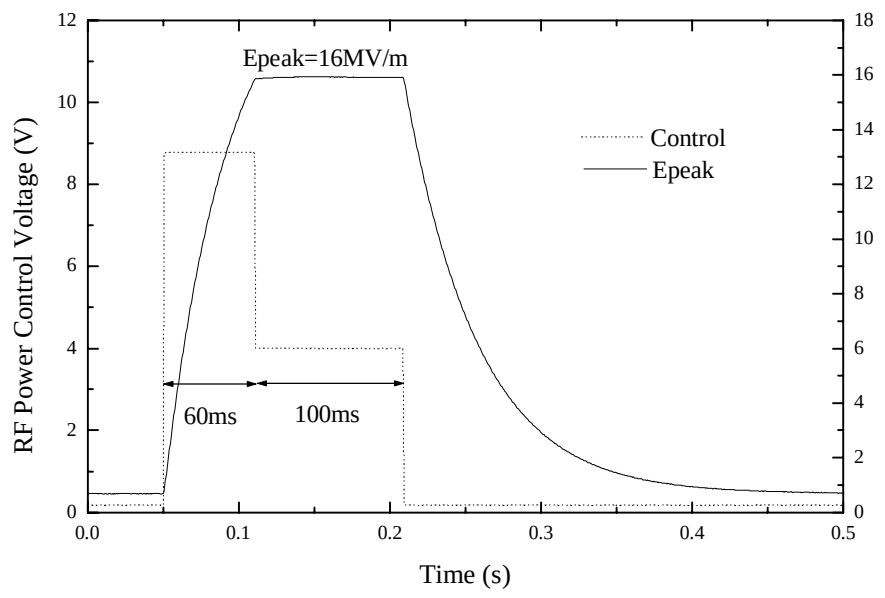

Fig. 1 Rf power control signal and Epeak in the pulsed operation

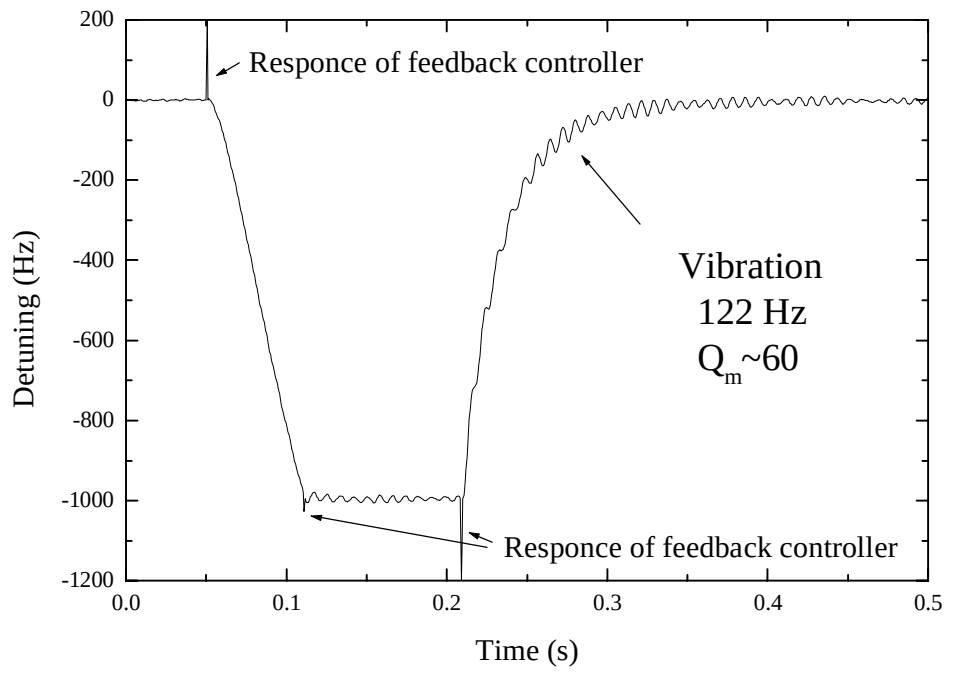

Fig. 2 Dynamic Lorentz Detuning obtained in the test

To prepare the parameters for the Lorentz Vibration Model, the SUPERFISH and the ABAQUS calculations were performed and then we found that only the first mode dominates the deformation. The frequency of the mode was calculated to be 111 Hz, which agreed well with the experimental results. In the Lorentz Vibration Model calculation, $V_C$ and $Q_m$ obtained experimentally were used. Figure 3 shows the calculated result compared with the experimental results. The average detuning at the flat top for the calculated and experimental data agreed within 10 %. The calculation also reproduces the behaviour of the vibration at the flat top and the decay. Since the geometry and the boundary conditions are very simple in this calculation, the agreement between the Model calculation and the experiment indicates the validity of the Lorentz Vibration Model.

Small disagreement between the measurement and the calculation shown in Fig. 3 is considered due to the errors of the parameters used in the calculation as well as measurement error. In applying the Lorentz Vibration Model, $\omega_{m1}$ and $K_1$ (only the first mode dominates the detuning in this case) were modified so as to reproduce the experimental data. Figure 4 shows the comparison of the modified calculated results and the experimental data at the flat top region. In the figure, the agreements between those data are very good.

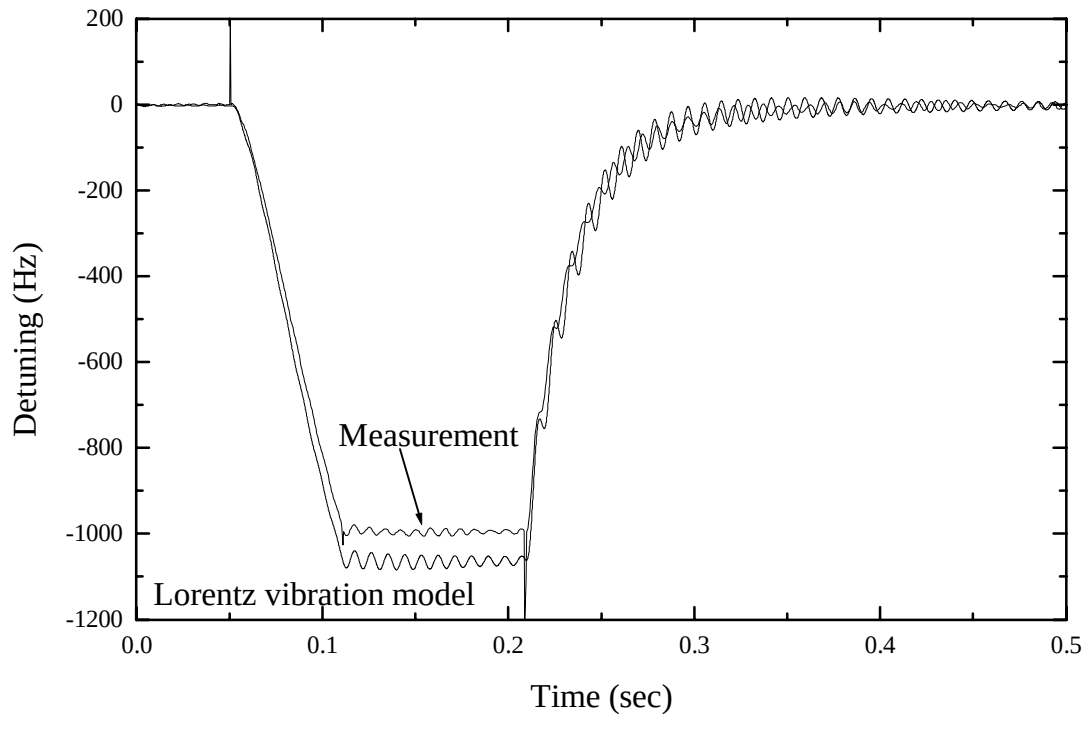

Fig. 3 Comparison between the calculated and the measured data

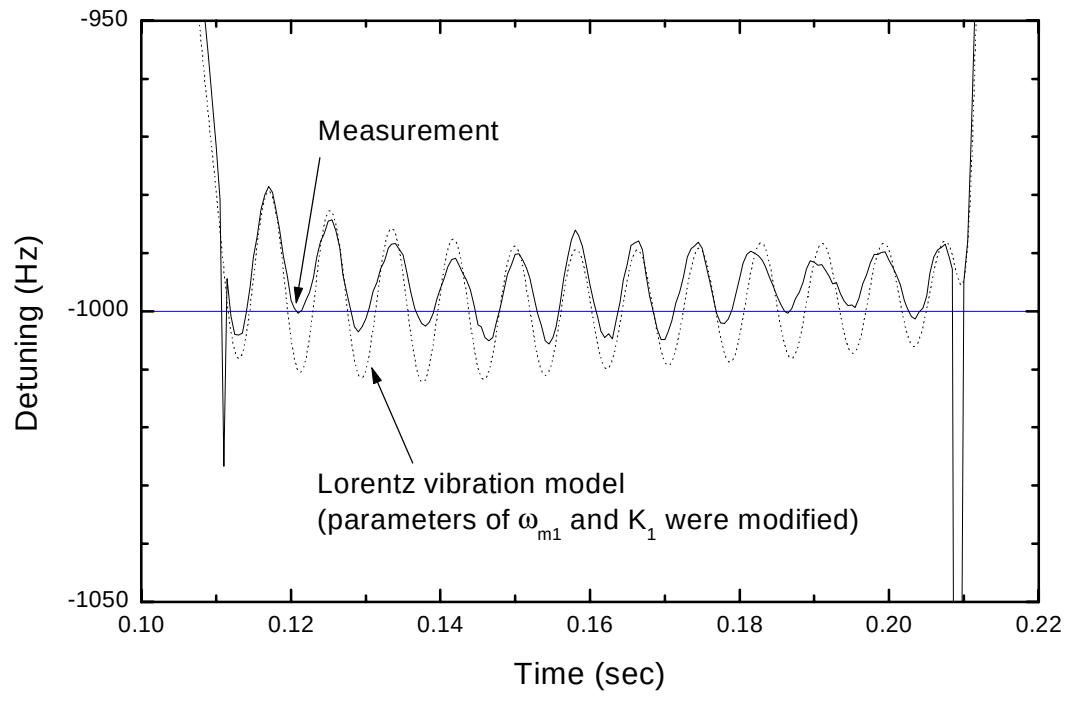

Fig. 4 Comparison between the modified calculation and the measured data at the flat top

# 3 DYNAMIC LORENTZ DETUNING FOR MULTI-CELL CAVITY

The Lorentz Vibration Model has been applied to the analysis of the dynamic detuning in the pulsed operation for the 972 MHz 7-cell cavity of $\beta_g$=0.729. The thickness of the cavity was set to be 2.8 mm.

## *3.1 Mechanical Modes of the 7-cell cavity*

At the first step of the analysis, 150 mechanical modes were calculated by the ABAQUS code. Figure 5 shows the typical modes as well as the stationary deformation by the Lorentz force. In this calculation, the left side of the cavity was fixed and the other side was supported by a spring as a tuner support. We found three kinds of mechanical modes; (a) multi-cell modes, in which modes, cell position moves with lower frequency, (b) tuner and beampipe modes, in which modes, only either end cell is deformed, and (c) single-cell mode, in which modes, cell position is fixed but each cell shape is deformed with higher frequency. Some of the single-cell modes have dominant influences to the detuning. Multi-cell modes have much less influences to the stationary detuning but are excited by the pulsed operation when the frequencies meet the multiple of the repetition rate.

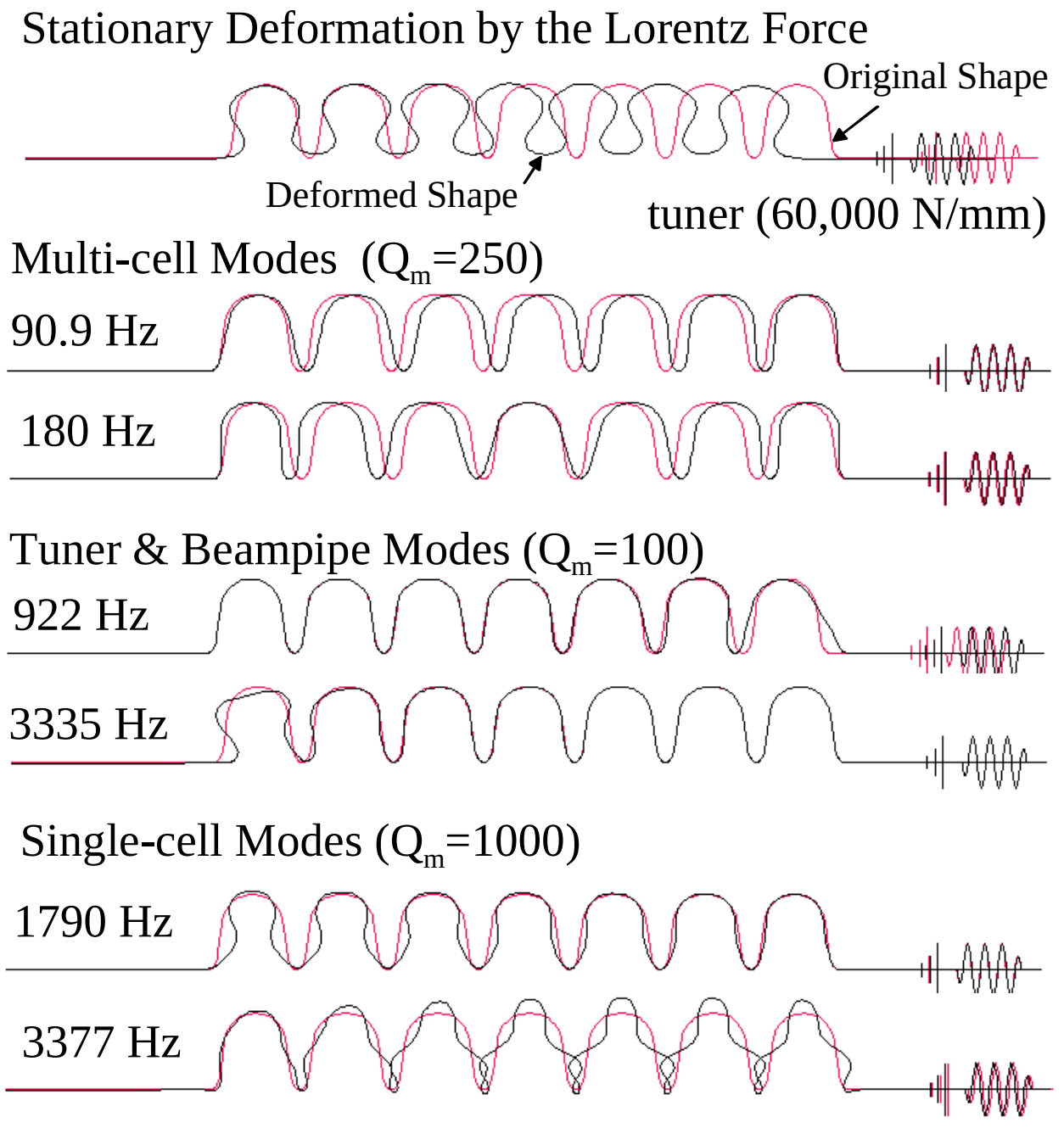

Fig. 5 Typical mechanical modes for the 972 MHz 7-cell cavity of $\beta_g$=0.729

Quality factors for the mechanical modes were set to be 250, 100 and 1000 for the multi-cell modes, tuner & beampipe modes and single-cell modes, respectively. Those values are based on our experimental experiences.

## *3.2 Choice of Mechanical Modes*

According to the Lorentz Vibration Model, we can consider the stationary condition by applying $(d^2\Delta f_k/dt^2)$ = $(d\Delta f_k/dt)$ = 0. Then we obtain the stationary detuning for each mode, $\Delta f_k=K_k(V_C/V_0)^2/\omega_{mk}^2$. From these data, we chose dominant 21 modes out of the 150 modes for the Lorentz Vibration Model calculation; 9 multi-cell modes, 2 tuner & beampipe modes and 10 single-cell modes.

## *3.3 Conventional Pulsed Operation*

The Lorentz Vibration Model was applied to the calculation of the dynamic detuning for the conventional pulsed operation, in which the cavity voltage increases exponentially for 0.6 ms, holds for 0.6 ms and decreases exponentially. Figure 6 shows the dynamic detuning, total detuning and some of the partial detuning, as well as the cavity voltage ($V_C$). In this calculation, some of the single-cell mode vibrations are excited by a pulsed voltage and the total detuning sways in the flat top and after the pulse. The vibration in the flat top causes cavity field error and the vibration after the pulse affects the next pulse.

## *3.4 Cosine-shaped Cavity Excitation*

In order to reduce the single-cell mode vibrations, we applied the cosine-shaped cavity excitation, in which cavity voltage increases and decreases in a cosine-shape. Figure 7 shows the cavity voltage and the dynamic detuning for the cosine-shaped cavity excitation. The rise time and the flat top duration were the same as the conventional pulsed operation. The cosine-shape decay was connected smoothly to the exponential decay. The vibration of the single-cell modes are much reduced as shown in Fig. 7.

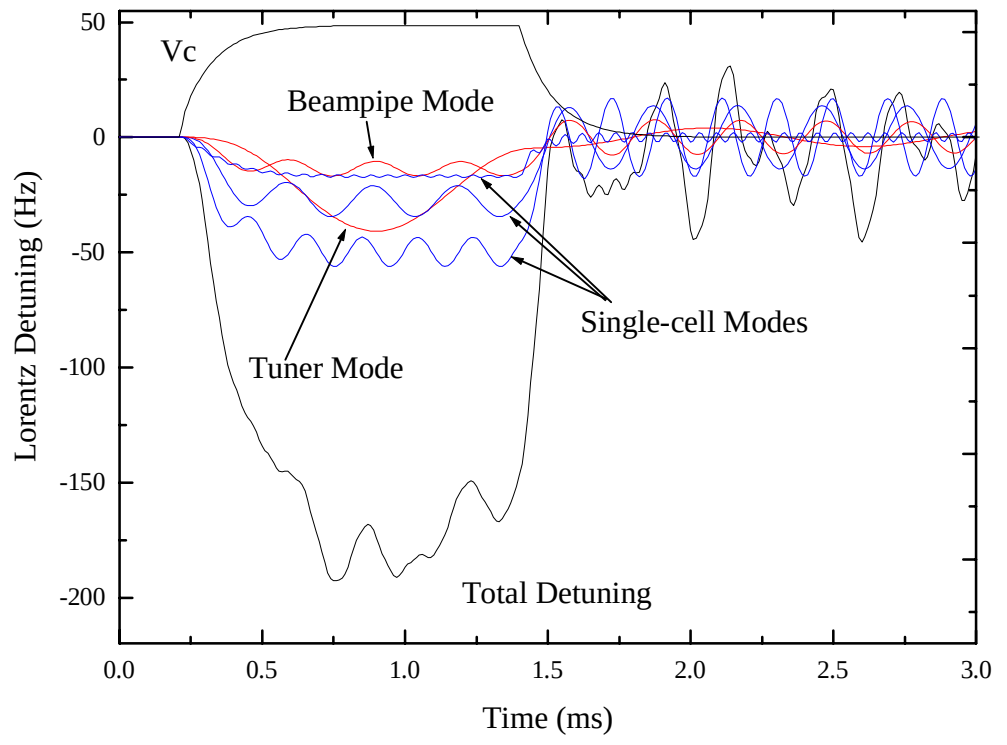

Fig. 6 Dynamic Lorentz detuning for the conventional pulsed operation

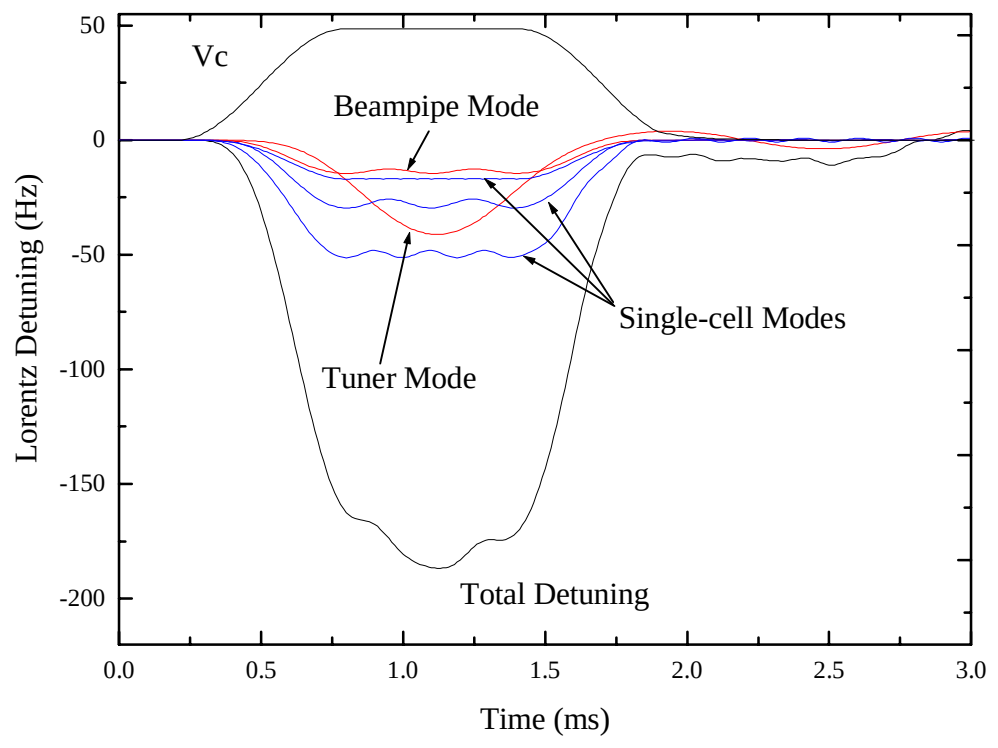

Fig. 7 Dynamic Lorentz detuning for the cosine-shaped cavity excitation

Since this method is considered to be suitable to obtain highly stable cavity field, it was applied to the rf control simulations described in the next section.

## 4 RF CONTROL SIMULATION

In the case that an rf source feeds the rf power to a single cavity, the good stability of the cavity field is expected because the influence of the dynamic Lorentz detuning can be compensated by an rf low level controller. In the system design of the JAERI/KEK Joint Project, two cavities in a cryomodule are controlled in one rf system. In this work, the rf control system of two cavities with individual mechanical properties, which is caused by fabrication errors, has been simulated. We assumed the different cavity wall thickness for providing individual mechanical properties and 2 simulations were performed; cavities of 2.8 mm and 3.2 mm thick, and 2.8mm and 3.0 mm thick. Loaded quality factor of the cavities are set to be $3.5\times10^5$ which is a half of the optimum one. This over-coupled condition moderates the influences of the dynamic detuning to the cavity field stability, even though additional rf power of about 20 % is required.

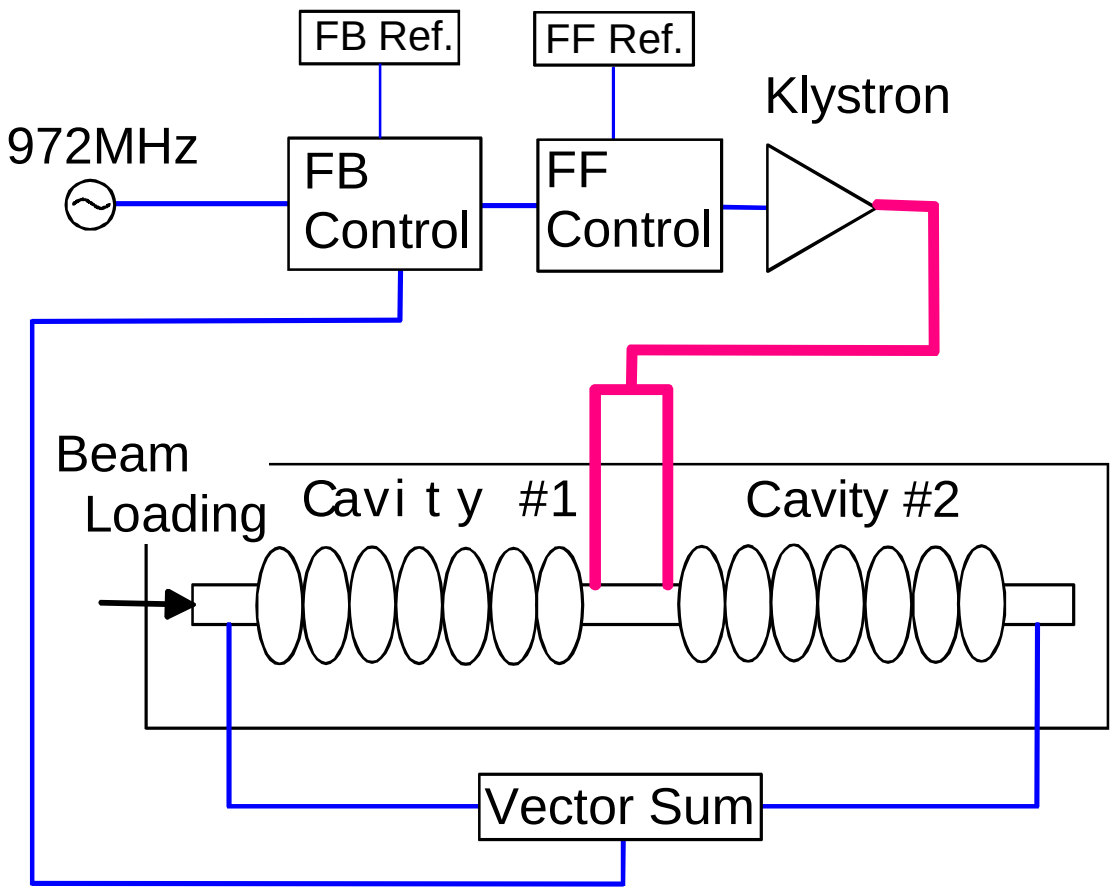

Fig. 8 Schematic block diagram of the rf control system

### 4.1 RF Control System

Figure 8 shows the schematic block diagram of the rf control system. In this simulation, vector sum control of two cavities were applied. The feed forward (FF) controller provides the cosine-shaped waveform and the waveform for the beam loading compensation. The cavity field stabilization against the dynamic detuning is performed by the feed back (FB) controller. The detune offset against the Lorentz detuning is optimized for each cavity because of the individual mechanical property.

### 4.2 Simulation with cavities of 2.8 mm and 3.2 mm thick

The rf control simulation with 2 cavities was carried out for 800 ms (40 pulses), where the wall thicknesses of cavity #1 and cavity #2 are 2.8 mm and 3.2 mm, respectively. The dynamic Lorentz detuning including the detune offset is plotted in Fig. 9 at every 0.1 second. In the figure, detuning of 0 degree means the optimum frequency. The proper offset for each cavity provides the good detuning in the beam period, ±~10Hz and ±~20Hz for the cavity #1 and #2, respectively. Figure 10 shows the amplitude and phase errors for those cavities. The amplitude errors up to ±~0.1 % and phase errors up to ±~0.2 deg were obtained for both cavities, while only the first pulse has slightly larger errors. Adopting the cosine-shaped cavity excitation and the proper detune offsets provide very good stability, which satisfies the requirement of ±1% and ±1deg.

### 4.3 Simulation with cavities of 2.8 mm and 3.0 mm thick

The rf control simulation with cavities of wall thicknesses of 2.8 mm (cavity #1) and 3.0 mm (cavity #3) were also carried out for 1000 ms (50 pulses). Unfortunately, the cavity #3 has a multi-cell mode of 349.5 Hz, which is very close to the multiple of the repetition rate, 350 Hz. Figure 11 shows the typical dynamic detuning for cavity #3. The oscillation of the multi-cell mode is emphasized significantly as shown in Fig. 11. The amplitude of the multi-cell mode is about 60 Hz. Figure 12 shows the dynamic Lorentz detuning including the detune offset. The detune offset for the cavity #3 (116 Hz) was provided with consideration of the excited multi-cell amplitude. Therefore, the detuning of the cavity #3 in the first pulse is far from the optimum position of 0 Hz in the beam period, but after several hundreds ms it becomes closer value as emphasizing the multi-cell mode. Figure 13 shows the amplitude and phase errors for cavity #1 and #3. The errors are large within several hundreds ms, however, they become smaller after that because of the stationary vibration of the multi-cell mode. At the time of 900 ms, we obtained the errors of ±0.15% and ±0.6deg for amplitude and phase, respectively. Even in this case, the stability of the cavity field satisfies the requirement.

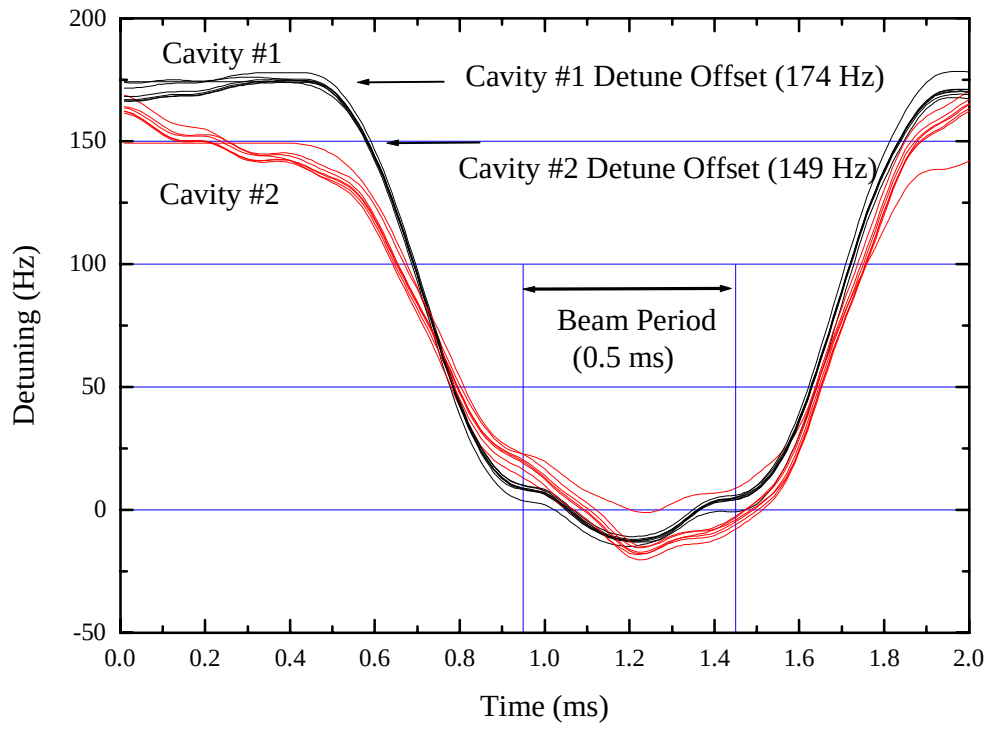

Fig. 9 Dynamic Lorentz detuning including the detune offset at every 0.1 s in the simulation for 2.8 mm and 3.2 mm thick cavities

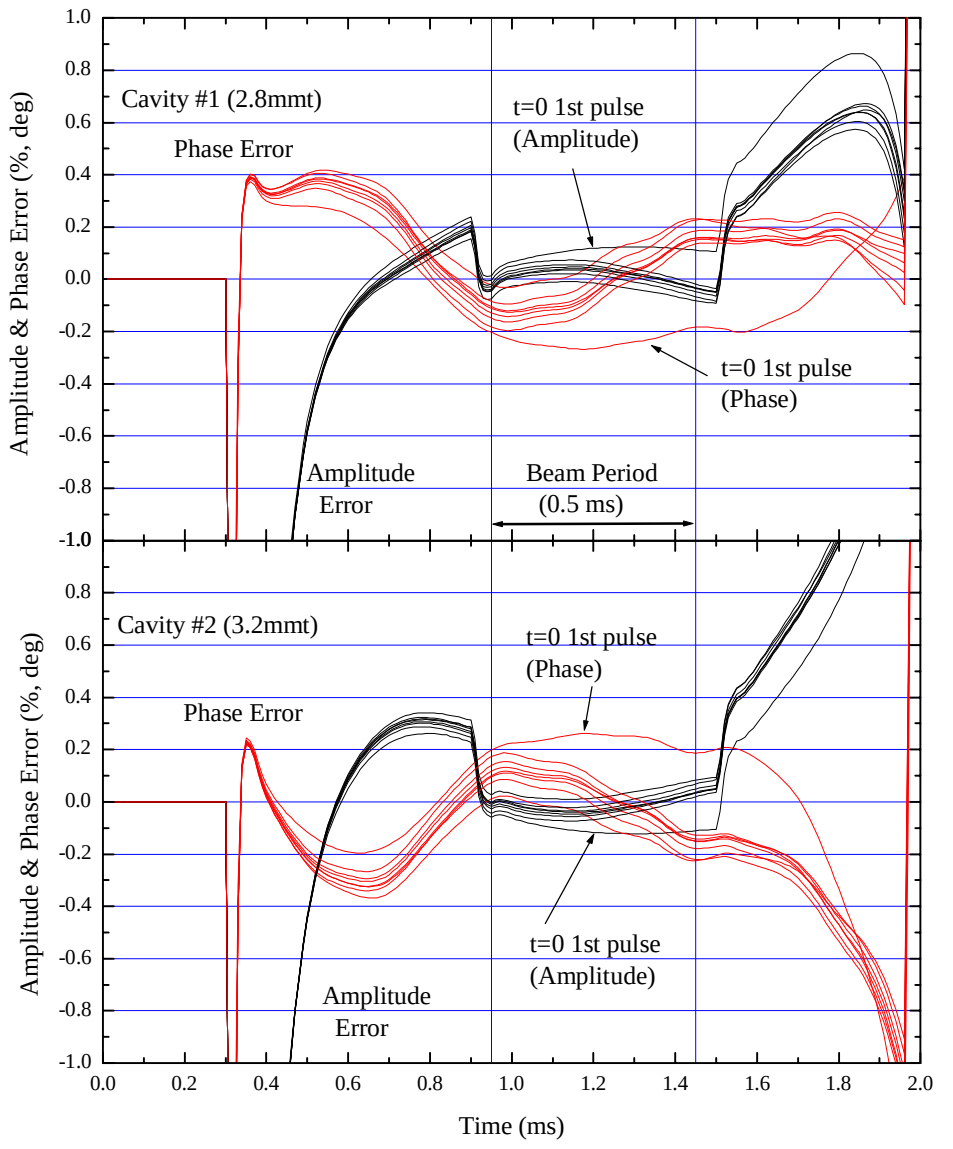

Fig. 10 Amplitude and phase errors at every 0.1 s in the simulation for 2.8 mm and 3.2 mm thick cavities

## 5 CONCLUSION

In order to simulate the rf control and to estimate the field stability of the SC proton linac, the Lorentz Vibration Model describing the dynamic Lorentz detuning has been developed. The validity of the new model has been confirmed experimentally. The model was applied successfully to the rf control simulation for 972 MHz 7-cell cavity of $\beta_g$=0.729. Here, we have obtained good cavity field stability which satisfied the requirement.

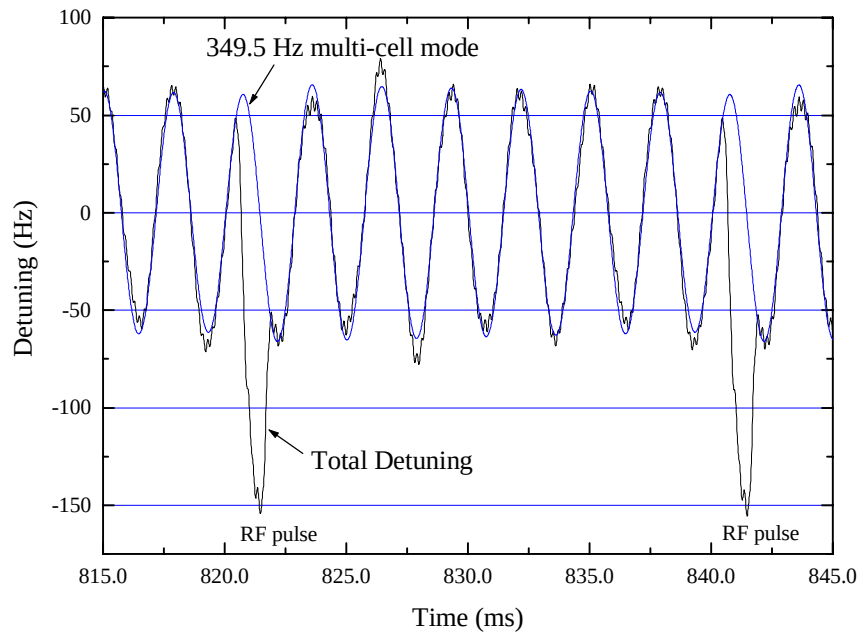

Fig. 11 Typical dynamic Lorentz detuning for the cavity #3 (3.0 mm thick)

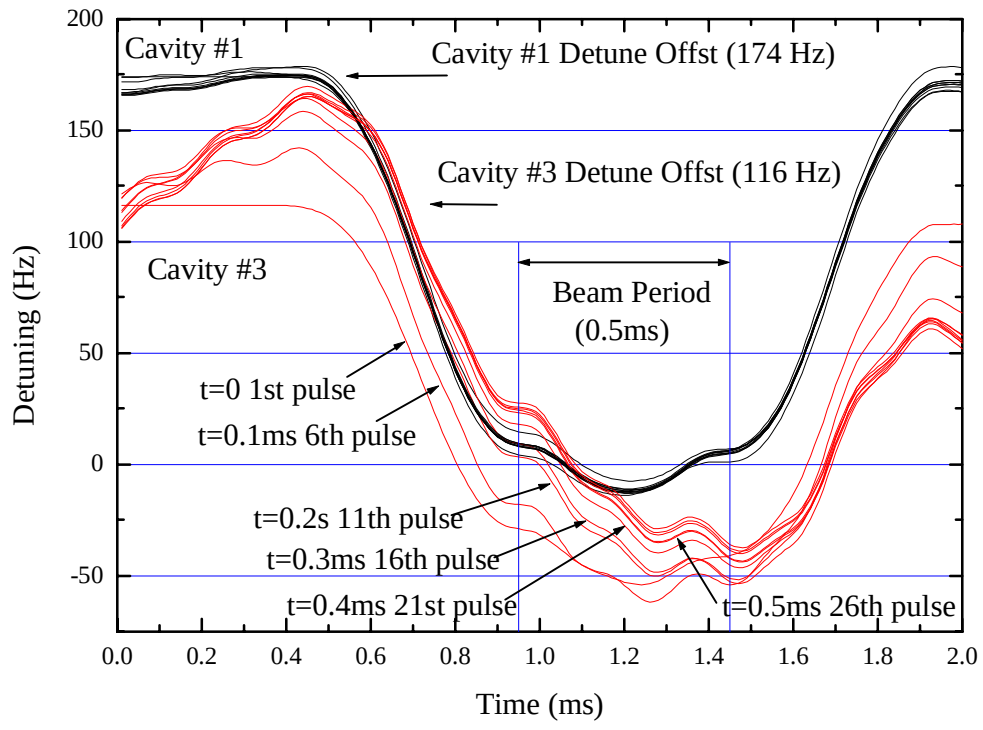

Fig. 12 Dynamic Lorentz detuning including the detune offset at every 0.1 s in the simulation for 2.8 mm and 3.0 mm thick cavities

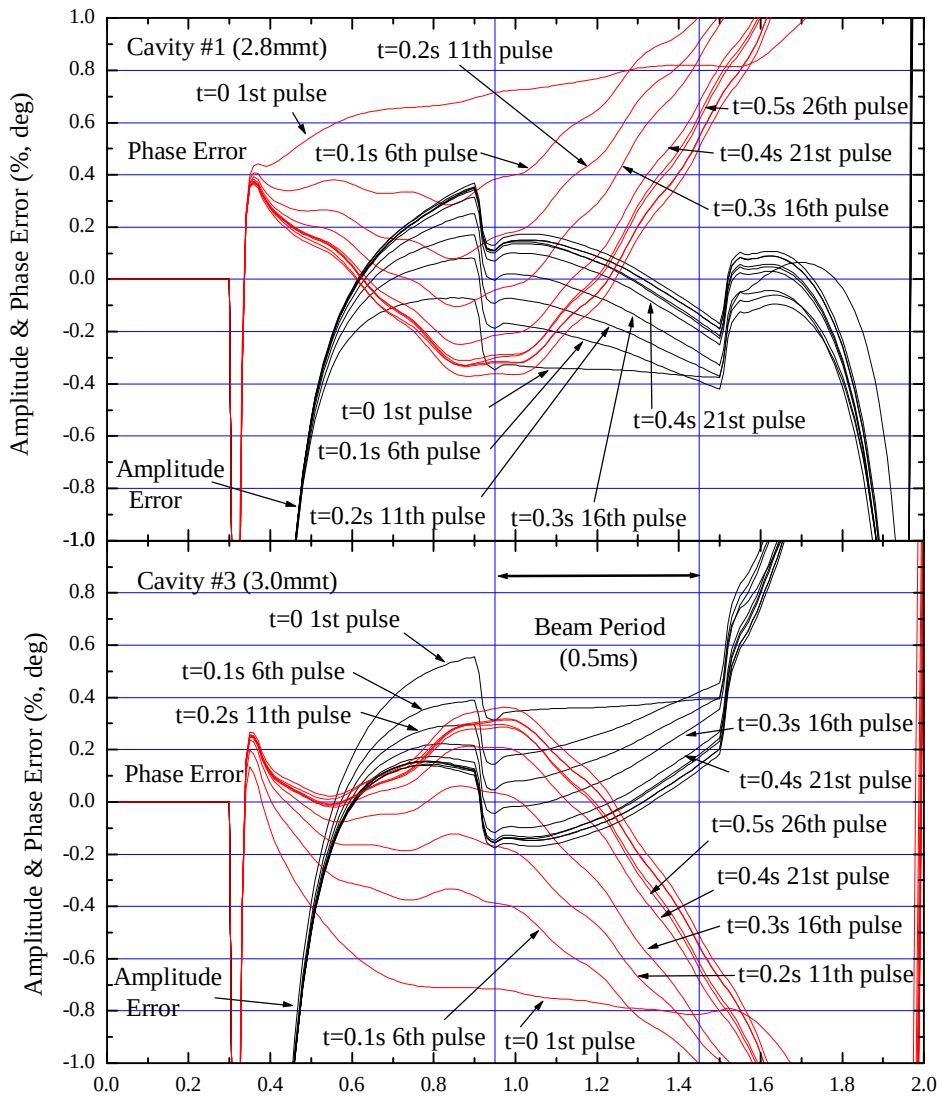

Fig. 13 Amplitude and phase errors at 0.1 s in the simulation for 2.8 mm and 3.0 mm thick cavities